\documentclass[aps,twocolumn,superscriptaddress,groupedaddress,preprintnumbers,floatfix,epfs]{revtex4}
\usepackage{amsfonts,amscd,amsmath,amssymb,graphicx,color,float}
\usepackage[section]{placeins}
\def\ep{\text{e}}
\def\g{\mathsf{g}}
\def\oh{\frac{1}{2}}
\def\s{\mathsf{s}}
\def\m{\mathsf{m}}
\def\k{\mathsf{k}}
\def\r{\mathsf{r}}
\def\rh{r_h}

\def\rq{r_q}
\def\rv{r_v}

\def\QQb{\text{\tiny Q}\bar{\text{\tiny Q}}}
\def\Qqb{\text{\tiny Q}\bar{\text{\tiny q}}}
\def\qQb{\text{\tiny q}\bar{\text{\tiny Q}}}
\def\Qqq{\text{\tiny Qqq}}
\def\Qqqb{\bar{\text{\tiny Q}}\bar{\text{\tiny q}}\bar{\text{\tiny q}}}

\begin{document}
\preprint{LMU-ASC 30/19}
\title{Baryon modes in string breaking from gauge/string duality}
\author{Oleg Andreev}
 \affiliation{L.D. Landau Institute for Theoretical Physics, Kosygina 2, 119334 Moscow, Russia}
\affiliation{Arnold Sommerfeld Center for Theoretical Physics, LMU-M\"unchen, Theresienstrasse 37, 80333 M\"unchen, Germany}
\begin{abstract} 
We consider the string breaking phenomenon within effective string models which purport to mimic QCD with two light flavors, with a special attention to baryon modes. We make some estimates of the string breaking distances at zero and non-zero baryon chemical potentials. Our estimates point towards the enhancement of baryon production in strong decays of heavy mesons in dense baryonic matter. We also suggest that the enhanced production of $\Lambda_c^+$ baryons in PbPb collisions is mainly due to larger values of chemical potential.
\end{abstract}
\maketitle
\textit{Introduction.--} The string models describe the strong decay of hadrons through light quark-antiquark pair creation \cite{strings}. A canonical example is the case in which a heavy meson decays into a pair of heavy-light mesons
\begin{equation}\label{mesonc}
	Q\bar Q\rightarrow Q\bar q+\bar Q q
	\,.
\end{equation}
In fact, this is one of the decay modes - the meson mode. One can think of it as a string rearrangement between the heavy and light sea quarks: $Q\bar Q+q\bar q\rightarrow Q\bar q+q\bar Q$. Despite the fact that in the vacuum the meson mode is dominant, it remains of interest to also consider other decay modes. The next, in number of light quarks, is the baryon mode 
\begin{equation}\label{baryonc}
	Q\bar Q\rightarrow Qqq+\bar Q\bar q\bar q
	\,,
\end{equation}
which is sub-dominant. Because it is not energetically favorable, and the probability for a string rearrangement between six quarks is lower than between four unless the sea quarks are regarded as a diquark-antidiquark pair $[qq][\bar q\bar q]$ \cite{diquarks}. In that case the string rearrangement occurs as before.

The presence of dense baryonic matter adds a crucial twist to the story, as light quarks are now not only due to pair creation but also to the medium. For this reason, a decay mode with three light quarks is allowed. It is  
\begin{equation}\label{hybrid}
	Q\bar Q\rightarrow Qqq+\bar Qq
	\,.
\end{equation}
Although it contains a meson, we will also call it a baryon mode because no confusion occurs with \eqref{baryonc} in what follows. It can be thought of as a string rearrangement between a diquark and quarks: $Q\bar Q+q[qq]\rightarrow Q[qq]+\bar Q q$, if a baryon is regarded as a two-body system composed of a quark and a diquark \cite{diquarks}.

In this letter we make some estimates of the string breaking distances and free energies of heavy-light hadrons using effective string models. Our goal is to gain a good intuition about nonperturbative phenomena in QCD in situations, where the use of other methods is impractical. 

\textit{The lattice model.--} The string breaking phenomenon arises from nonperturbative effects of QCD. So far lattice QCD has successfully been used to study it, but only for meson modes at zero temperature and zero chemical potential. For our purposes, what we need to know can be summarized as follows. The model of \cite{drum} includes a mixing analysis based on a correlation matrix whose elements give rise to a model Hamiltonian. The eigenvalues of this Hamiltonian correspond to the energy levels. In this way, the lattice data are well-described by a few fit parameters \cite{bali,bulava}.

Certainly, this model can be extended in several ways, for instance, by adding new light flavors, by considering baryon modes, or by turning on chemical potential. We will consider only two cases. In the first case, modes \eqref{mesonc} and \eqref{baryonc}, the model Hamiltonian is 
\begin{equation}\label{Vmatrix}
H(\ell)=
\begin{pmatrix}
E_{\QQb}(\ell) & g_1 & g_2\\
g_1& 2E_{\Qqb} &g_{12}\\
g_2 & g_{12} & 2E_{\Qqq}	
\end{pmatrix}
\,,
\end{equation}
where $E_{\QQb}(\ell)$ is the energy of two static heavy quark sources separated by distance $\ell$ \cite{V0}. $2E_{\Qqb}$ and $2E_{\Qqq}$ are the energies of a noninteracting pair of heavy-light mesons and baryons, respectively. The off-diagonal matrix elements $g$ describe the mixing between the states. A useful parameter, called the string breaking distance, is defined by setting $E_{\QQb}(\ell_c^{(m)})=2E_{\Qqb}$ and $E_{\QQb}(\ell_c^{(b)})=2E_{\Qqq}$ for the meson and baryon modes, respectively \cite{drum,bulava}. 

In the second case, modes \eqref{mesonc} and \eqref{hybrid}, the free energy matrix is given by 
\begin{equation}\label{Vmatrix-mu}
F(\ell,\mu)=
\begin{pmatrix}
F_{\QQb}(\ell) & f_1 & f_2\\
f_1& F_{\Qqb}+F_{\qQb} &f_{12}\\
f_2 & f_{12} & F_{\Qqq}+F_{\qQb}	
\end{pmatrix}
\,,
\end{equation}
where $F_{\QQb}(\ell)$ is the free energy of two heavy quark sources in the medium. $F_{\Qqb}$, $F_{\qQb}$ and $F_{\Qqq}$ are the free energies of noninteracting  heavy-light mesons and baryons. As before, the off-diagonal matrix elements describe the mixing between the states. By analogy with what was done for the Hamiltonian \eqref{Vmatrix}, we now define the string breaking distances by setting $F_{\QQb}(\ell_c^{(m)})=F_{\Qqb}+F_{\qQb}$ and $F_{\QQb}(\ell_c^{({b'})})=F_{\Qqq}+F_{\qQb}$.

\textit{String Models.--} Although lattice QCD is a powerful computational tool to deal with strongly coupled gauge theories, its use is limited when it comes to studying baryonic matter whose density is not really small. Meanwhile the gauge/string duality allows one to at least get a good intuition in such situations. It is noteworthy that the duality was first proposed for conformal theories and then expanded to include non-conformal ones, and thus apply these ideas to QCD and heavy ion collisions \cite{uaw-book}. 

In this formalism a Wilson loop ${\cal C}$ is placed on an Euclidean four-manifold which is the boundary of a five-dimensional manifold. Its expectation value is given by the world-sheet path integral so that a string world-sheet has ${\cal C}$ for its boundary.  In principle, the integral can be evaluated semiclassically. The result is written as 
\begin{equation}\label{wilson}
\langle\,W({\cal C})\,\rangle=\sum_n w_n\ep^{-S_n}
\,,
\end{equation}
where $S_n$ is the world-sheet action evaluated on a classical solution (string configuration). $n$ labels the solutions. $w_n$ is a relative weight factor. For static configurations each $S_n$ reduces to $E_n{\cal T}$, where $E_n$ is an energy of the configuration and ${\cal T}$ is a time interval. Importantly, the $E_n$'s are the diagonal elements of the lattice model Hamiltonian. So, on the left hand side of \eqref{wilson}, two exponents become equal at the string breaking distance. 

The generalization of this approach to the case of finite temperature and baryon density suggests that the correlator of two oppositely oriented Polyakov loops is given by the worldsheet path integral, and it can be evaluated again as 
\begin{equation}\label{polyakov}
\langle\,L\,L^{\dagger}\,\rangle=\sum_n \omega_n\ep^{-S_n}
\,. 
\end{equation}
For static configurations each $S_n$ reduces to $F_n/T$, where $F_n$ is a free energy of the configuration and $T$ is temperature.

To construct string configurations, one needs two basic objects. The first is a Nambu-Goto string governed by the action 
\begin{equation}\label{NG}
S_{\text{\tiny NG}}=\frac{1}{2\pi\alpha'}\int d^2\xi\,\sqrt{\gamma^{(2)}}
\,,
\end{equation}
with $\gamma$ an induced metric, $\alpha'$ a string parameter, and $\xi$ worldsheet coordinates. The second is a baryon vertex. It is a fivebrane in ten dimensions \cite{witten}, but a point-like object from the five-dimensional viewpoint. At leading order in $\alpha'$, the brane dynamics is determined by its world volume.

\textit{Zero temperature and chemical potential.--} When we try to mimic QCD with $N_c=3$ and $N_f=2$, we consider an effective string model on five-dimensional space, with the Euclidean metric 
\begin{equation}\label{metric}
ds^2=\ep^{\s r^2}\frac{R^2}{r^2}
\Bigl(dt^2+dx_i^2+dr^2\Bigr)
\,,
\end{equation}
times a five-dimensional compact space $X$. For simplicity, we assume no dependence on coordinates of $X$ and therefore the model under consideration is in fact five-dimensional. This geometry is a one-parameter deformation, parameterized by $\s$ \cite{slant}, of the Euclidean $\text{AdS}_5$ space of radius $R$ whose boundary is at $r=0$. Its main feature is a soft wall located at $r=1/\sqrt{\s}$. There are good motivations for taking it seriously. First, such a deformation leads to linear Regge-like spectra for the light mesons \cite{son}. Second, it gives rise to the heavy quark-antiquark potential which is very satisfactory in the light of lattice gauge theory and phenomenology \cite{az1,giannuzzi}. 

In addition, we add an open string tachyon background \cite{tach}. There are at least two good reasons for this. First, a constant tachyon field can be interpreted as a point-like mass in five dimensions so that a string can terminate on it in the bulk. This is not a new idea and many aspects of strings with masses at the ends have been described long ago in flat space \cite{bn-book}. The novelty is that it is used in the context of the gauge/string duality to model light flavors. Second, a non-constant field can carry information on the chiral condensate \cite{son} that opens a promising avenue of possible studies aimed to gain insight into non-perturbative aspects of QCD, in particular its phase diagram. For the reason of simplicity, we consider a constant tachyon background. In this case, the action is simply
\begin{equation}\label{quark}
S_{\text{q}}=\tau_0\int ds
\,,
\end{equation}
where $\tau_0$ is a mass parameter. The integral is carried out over a worldsheet boundary.

\begin{figure}[ht]
\centering
\includegraphics[width=7.5cm]{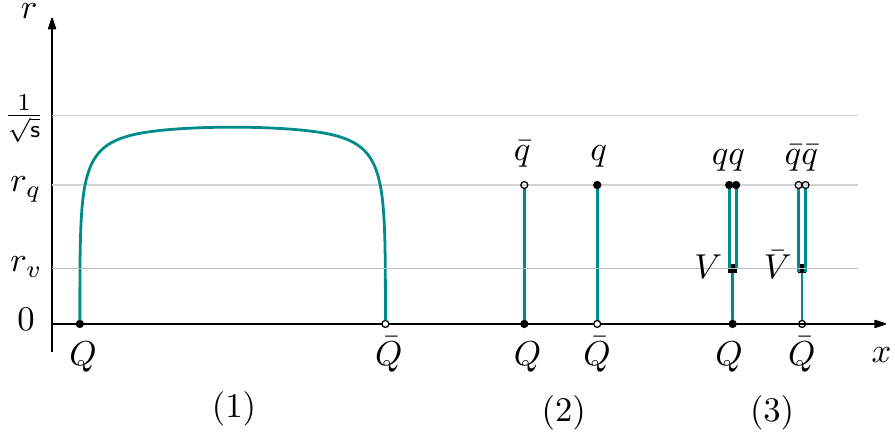}
\caption{{\small Static string configurations arranged in a number of light quarks. The boundary is at $r = 0$. The heavy (light) quarks are denoted by $Q$ ($q$), and the baryon vertices by $V$.}}
\label{conf0}
\end{figure}

Now consider a rectangular Wilson loop of size $\ell\times{\cal T}$ such that ${\cal T}\gg\ell$. The dominant string configurations, for large $\ell$ are shown in Figure \ref{conf0}. They 
are the string duals to the three states of the lattice model. In this picture a diquark looks like a one-dimensional object extended along the fifth dimension and constructed from two light quarks and a baryon vertex.  

Since we are interested in static configurations, we choose the static gauge. Then the analysis of configuration (1) associated to a single Nambu-Goto string proceeds just as in \cite{az1}. At large distances, the energy of the pair is given by \cite{largeL}
\begin{equation}\label{EQQb}
E_{\QQb}(\ell)=\sigma\ell-2\g\sqrt{\s}\,I_0+C+o(1)
\,,
\end{equation}
where
\begin{equation}\label{I0}
\sigma=\ep\g\s
\,,\quad
I_0=\int_0^1\frac{dx}{x^2}\biggl(1+x^2-\ep^{x^2}\Bigl[1-x^4\ep^{2(1-x^2)}\Bigr]^{\frac{1}{2}}\biggr)
\,,
\end{equation}
$\g=\frac{R^2}{2\pi\alpha'}$, and $C$ is a renormalization constant. Numerically, $I_0\approx 0.751$. 

In the case of configuration (2), the action has in addition to the standard Nambu-Goto actions of the fundamental strings, contributions arising from the light quarks. It is thus $S=\sum_{i=1}^2 S_{\text{\tiny NG}}+S_{\text{q}}$. Varying $S$ with respect to $r_q$ yields \cite{strb}
\begin{equation}\label{fb-q}
\g\,\ep^{\frac{q}{2}}+\m(q-1)=0
\,,
\end{equation}
with $\m=\tau_0R$ and $q=\s\rq^2$. It describes the force balance condition in the radial direction for the light quark. At zero chemical potential it has the same form for both the quark and the antiquark. Then a simple analysis shows that the energy of the configuration is 
\begin{equation}\label{EQqb}
E_{\Qqb}+E_{\qQb}=2\sqrt{\s}\,{\cal Q}_0(\g,\m,q)+C
\,,
\end{equation}
where the function ${\cal Q}_0$ is defined in the Appendix.  

A similar analysis for configuration (3) would proceed in essentially the same way. The only novelty is the baryon vertex which in static gauge is governed by the action 
\begin{equation}\label{baryon-v}
S_{\text{vert}}=\tau_v\frac{\ep^{-2\s r^2}}{r}{\cal T}
\,,
\end{equation}
with $\tau_v$ a dimensionless parameter. Such a form comes from the brane world volume and leads to the very satisfactory description of the lattice data obtained for the three-quark potential \cite{a-3q}. Given this, we can write for the total action of the configuration $S=\sum_{i=1}^2 3S_{\text{\tiny NG}}+2S_{\text{q}}+S_{\text{vert}}$. Varying the action with respect to $r_q$ gives again Eq.\eqref{fb-q}, while that with respect to $r_v$ gives \cite{strb}
\begin{equation}\label{fb-v}
1+3\k(1+4v)\ep^{-3v}=0
\,.
\end{equation}
Here $v=\s\rv^2$ and $\k=\frac{\tau_v}{3\g}$. This equation describes the force balance condition in the radial direction for the vertex. Further, it can be shown \cite{strb} that the energy of the configuration is 
\begin{equation}\label{EQqq}
E_{\Qqq}+E_{\Qqqb}=
2\sqrt{\s}
\Bigl(2{\cal Q}_0(\g,\m,q)+{\cal V}_0(\g,v)\Bigr)
+C
\,,
\end{equation}
where the function ${\cal V}_0$ is given by \eqref{V}. Note that this configuration exists only if $q>v$. 

Given the energies of the three states, it is straightforward to write down the expressions for the string breaking distances. So we have
\begin{gather}
\ell_c^{(m)}=\frac{2}{\ep\g\sqrt{\s}}
\Bigl({\cal Q}_0(\g,\m,q)+\g I_0\Bigr)
\,,
\label{lc0}
\\
\ell_c^{(b)}=\frac{2}{\ep\g\sqrt{\s}}
\Bigl(
2{\cal Q}_0(\g,\m,q)+{\cal V}_0(\g,v)+\g I_0
\Bigr)
\,.
\label{lcd0}
\end{gather}
Importantly, the dependence on $C$ cancels, and as a result the $\ell$'s are scheme-independent. 

To make estimates, we will need values of the fit parameters. There are two ways to choose those values: one is to use the lattice QCD results, and the other is to follow closely phenomenology. In the first way, the value of $\s$ is fixed from the slope of the Regge trajectory of $\rho(n)$ mesons in the soft wall model with the geometry \eqref{metric}. This gives $\s=0.45\,\text{GeV}^2$ \cite{a-q2}. Then, using \eqref{I0}, we obtain $\g=0.176$ by fitting the string tension to its value in \cite{bulava}. Next, the parameter $\m$ is adjusted to reproduce the lattice result for the string breaking distance $\ell_c^{(m)}$. With $\ell_c^{(m)}=1.22\,\text{fm}$ \cite{bulava}, it gives $\m=0.538$. Using Eq.\eqref{fb-q}, $q$ can be estimated to be about $0.566$. A simple analysis shows that on the interval $[0,0.566]$ Eq.\eqref{fb-v} has solutions if $-0.558\lesssim \k\leq -\tfrac{1}{4}\ep^{\frac{1}{4}}$. In the string models \cite{a-3q}, the value of $\k$ is adjusted to fit the three-quark potential to the lattice data for pure $SU(3)$ gauge theory. So far there is no such data available for QCD with dynamical quarks. We take $\k=-\tfrac{1}{4}\ep^{\frac{1}{4}}\approx-0.321$ simply because it yields an exact solution to Eq.\eqref{fb-v}, namely $v=\tfrac{1}{12}$. We will denote this set of values by $L$. With $L$, we get
\begin{equation}\label{lcb-lattice}
\ell^{(b)}_c=2.35\,\text{fm}
\,.
\end{equation}
So the baryon mode is sub-dominant, as expected. Importantly, in the above interval for $\k$ the breaking distance $\ell_c^{(b)}$ is a slowly varying function of $\k$ which can take values from $2.23\,\text{fm}$ to $2.35\,\text{fm}$. Thus, in \eqref{lcb-lattice} the error related with our choice of $\k=-\tfrac{1}{4}\ep^{\frac{1}{4}}$ does not exceed $5\%$.

The lattice calculation was done at unphysical pion mass $m_{\pi}=280\,\text{MeV}$ \cite{bulava}. In this light and in view of possible applications to phenomenology, we now make the estimates for other values of the fit parameters. First, the values of $\s$ and $\g$ are extracted from the quarkonium spectrum obtained by using the heavy quark potential derived from the model we are considering \cite{az1}. This is self-consistent, and gives $\s=0.15\,\text{GeV}^2$ and $\g=0.44$ \cite{giannuzzi}. We set $\k=-\tfrac{1}{4}\ep^{\frac{1}{4}}$, as before. Then we determine $\m$ from the condition $E_{\Qqq}-E_{\Qqb}=M_{\Lambda_c^+}-M_{D^0}\approx 420\,\text{MeV}$ \cite{pdg}. It results in $\m=0.699$. We will denote this set of values by $P$. For $P$, 
\begin{equation}\label{lc=pheno}
\ell_c^{(m)}=1.07\,\text{fm}\,,
\qquad
\ell_c^{(b)}=1.99\,\text{fm}
\,.
\end{equation}
These are smaller than the previous ones that could correspond to a more physical situation, with a lighter pion.

\textit{Cold baryonic matter.--} Now we extend the above analysis to include baryonic matter at zero temperature. We do this for an effective string model which is the simplest extension of the above model to non-zero baryon chemical potential. A key point is that the background geometry is now given by a one-parameter deformation of the Reissner-Nordstr\"om charged black hole in Euclidean $\text{AdS}_5$ \cite{PC1}. In the case of interest, it is given by 
\begin{equation}\label{RNs}
ds^2=\ep^{\s r^2}\frac{R^2}{r^2}
\Bigl(fdt^2+dx_i^2+\frac{dr^2}{f}\Bigr)
\,,\quad
A_0(r)=\mu-\r\frac{r^2}{\rh^3}
\,,
\end{equation}
with $f(r)=\bigl(1-\bigl(\frac{r}{\rh}\bigr)^2\bigr)^2\bigl(1+2\bigl(\frac{r}{\rh}\bigr)^2\bigr)$. The black hole horizon is at $r=\rh$. $\mu$ denotes a baryon chemical potential, and $\r$ a free parameter. The gauge field vanishes at the horizon that gives a relation between $\mu$ and $\rh$. It is convenient to write it as $\mu=\r\sqrt{\frac{\s}{h}}$, with $h=\s\rh^2$. The model is in the confined phase for $h>h_c$, where $h_c\approx 3.50$ \cite{strb}. Therefore the maximal value of $\mu$ is given by $\mu_{pc}=\r\sqrt{\frac{\s}{h_c}}$.

As usual, in the presence of the background gauge field the world-sheet action includes a term which, in our case, reduces to 
\begin{equation}\label{SA}
S_{\text{\tiny A}}=\mp\frac{1}{3}\int dt\, A_0
\,.
\end{equation} 
Here the minus and plus signs correspond to a quark and an antiquark, respectively. The prefactor comes from the relation between the chemical potentials of quarks and baryons.

\begin{figure}[ht]
\includegraphics[width=7.5cm]{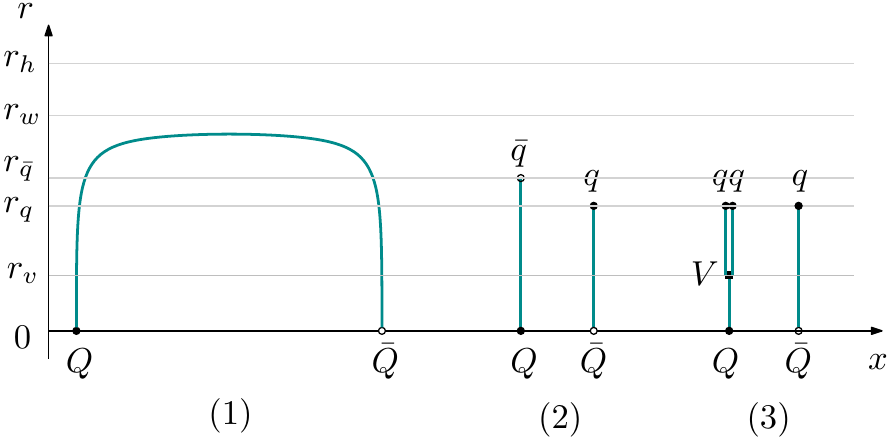}
\caption{{\small Static string configurations with the light quark number up to $3$. The soft wall is at $r=r_w$ so that $r_h>r_w$. Because the potential $A_0$ gives the electric field in the $r$-direction, the light quarks are shifted towards the boundary.}}
\label{confm} 
\end{figure}

Now let us consider the correlator of two Polyakov loops in the low-temperature limit. The dominant string configurations in this limit are shown in Figure \ref{confm}. These are the 
string duals to the three states of the lattice model. In the static gauge, these configurations can be analyzed in a way which is similar to that above. 

For configuration (1), we find that at large distances the free energy of the pair takes the form \cite{strb}
\begin{equation}\label{FQQb}
F_{\QQb}(\ell)=\sigma(\mu)\ell-2\g\sqrt{\s}\,I+C+o(1)
\,,
\end{equation}
with $\sigma(\mu)$ and $I$ given by \eqref{I}. At $\mu=0$, $I=I_0$.

Configuration (2) can be analyzed as before. When one adds the $S_{\text A}$'s to the total action, one finds that Eq.\eqref{fb-q} becomes 
\begin{equation}\label{fb-qmu}
\g\ep^{q}
+
\m
\sqrt{1+2\tfrac{q}{h}}
\biggl((q-1)\bigl(1-\tfrac{q}{h}\bigr)-\frac{6\bigl(\frac{q}{h}\bigr)^2}{1+2\frac{q}{h}}\biggr)\ep^{\frac{q}{2}}
=\mp
\frac{2}{3}\r\bigl(\tfrac{q}{h}\bigr)^{\frac{3}{2}}
\,.
\end{equation}
Here the minus sign refers to quarks and the plus sign to antiquarks, where $q$ is replaced by $\bar q=\s r^2_{\bar q}$. The free energy of the configuration is given by \cite{strb}
\begin{equation}\label{FQqb}
F_{\Qqb}+F_{\qQb}=\sqrt{\s}\Bigl({\cal Q}\bigl(\g,\m,-\r,\bar q,\tfrac{\bar q}{h}\bigr)+{\cal Q}\bigl(\g,\m,\r,q,\tfrac{q}{h}\bigr)\Bigr)+C
\,,
\end{equation}
where the function ${\cal Q}$ is defined in the Appendix. 

It is also straightforward to generalize the above analysis 
to configuration (3). In this case, the action for the baryon vertex becomes 
\begin{equation}\label{baryon-vmu}
S_{\text{vert}}=\tau_v\sqrt{f}\frac{\ep^{-2\s r^2}}{r}T^{-1}
\,.
\end{equation}
The total action is amended by adding the $S_{\text A}$'s. Varying with respect to $r_q$ gives again Eq.\eqref{fb-qmu}, and that with respect to $r_v$ gives  
\begin{equation}\label{fb-vmu}
\sqrt{1+2\tfrac{v}{h}}+3\k
\Bigl((1+4v)\Bigl(1+\tfrac{v}{h}-2\bigl(\tfrac{v}{h}\bigr)^2\Bigr)
+6\bigl(\tfrac{v}{h}\bigr)^2
\Bigr)\ep^{-3v}=0
\,.
\end{equation}
As shown in \cite{strb}, the free energy of this configuration is 
\begin{equation}\label{FQqq}
F_{\Qqq}+F_{\qQb}=\sqrt{\s}\Bigl(
3{\cal Q}\bigl(\g,\m,\r,q,\tfrac{q}{h}\bigr)+{\cal V}\bigl(\g,v,\tfrac{v}{h}\bigr)
\Bigr)
-\mu+C
\,,
\end{equation}
where the function ${\cal V}$ is given by \eqref{V}. 

If one is given the free energy of all the three states, then one can easily find the corresponding string breaking distances 
\begin{gather}
\ell_c^{(m)}=\frac{\sqrt{\s}}{\sigma(\mu)}
\Bigl(
{\cal Q}\bigl(\g,\m,\r,q,\tfrac{q}{h}\bigr)+{\cal Q}\bigl(\g,\m,-\r,\bar q,\tfrac{\bar q}{h}\bigr)
+2\g I
\Bigr)
\,,
\label{lc-mu}
\\
\ell_c^{(b')}=\frac{\sqrt{\s}}{\sigma(\mu)}
\Bigl(
3{\cal Q}\bigl(\g,\m,\r,q,\tfrac{q}{h}\bigr)+{\cal V}\bigl(\g,v,\tfrac{v}{h}\bigr)
-\tfrac{\mu}{\sqrt{\s}}
+2\g I
\Bigr)
.
\label{lcmb-mu}
\end{gather}

\begin{figure*}[htbp]
\centering
\includegraphics[width=7cm]{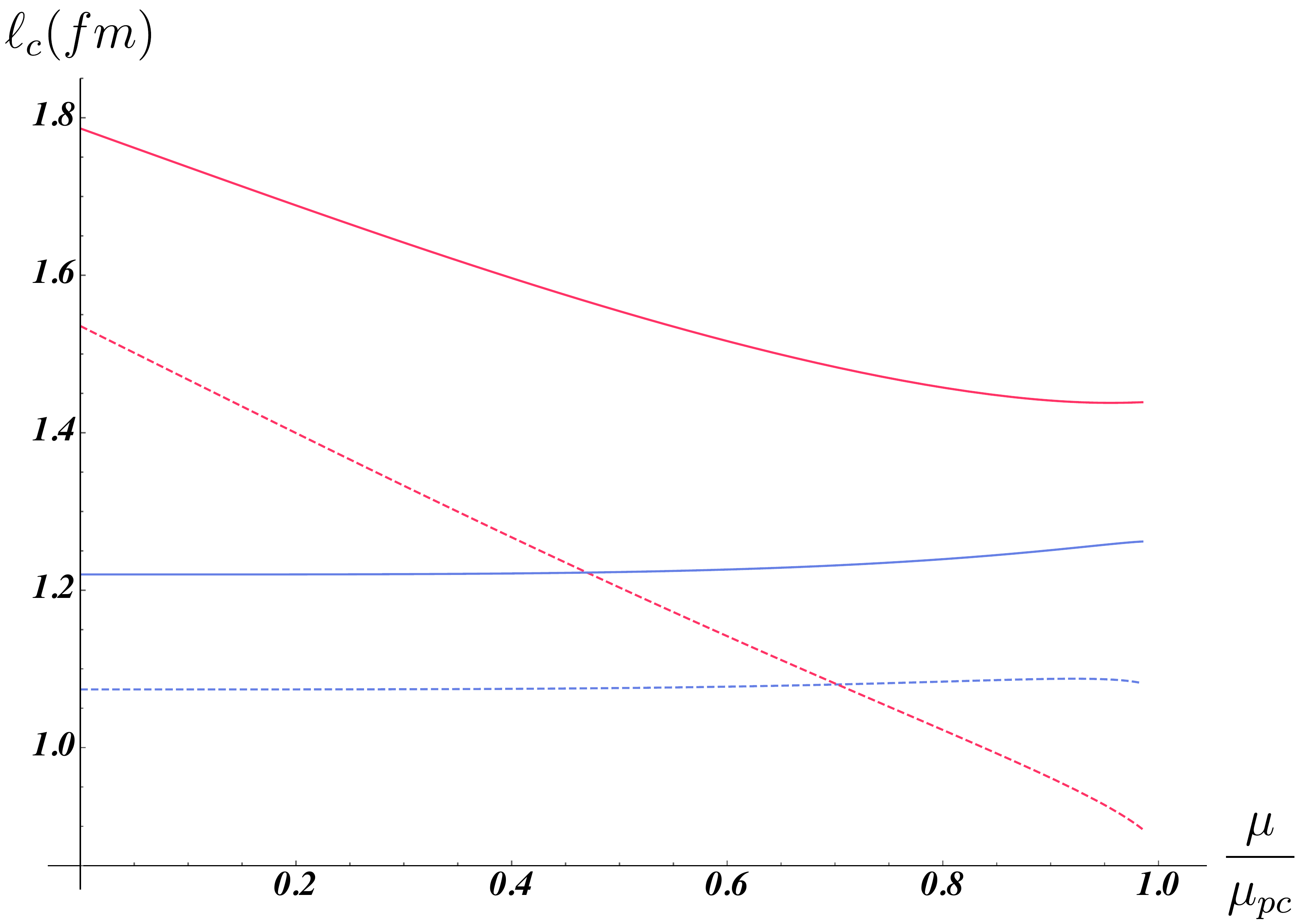}
\hspace{2cm}
\includegraphics[width=7cm]{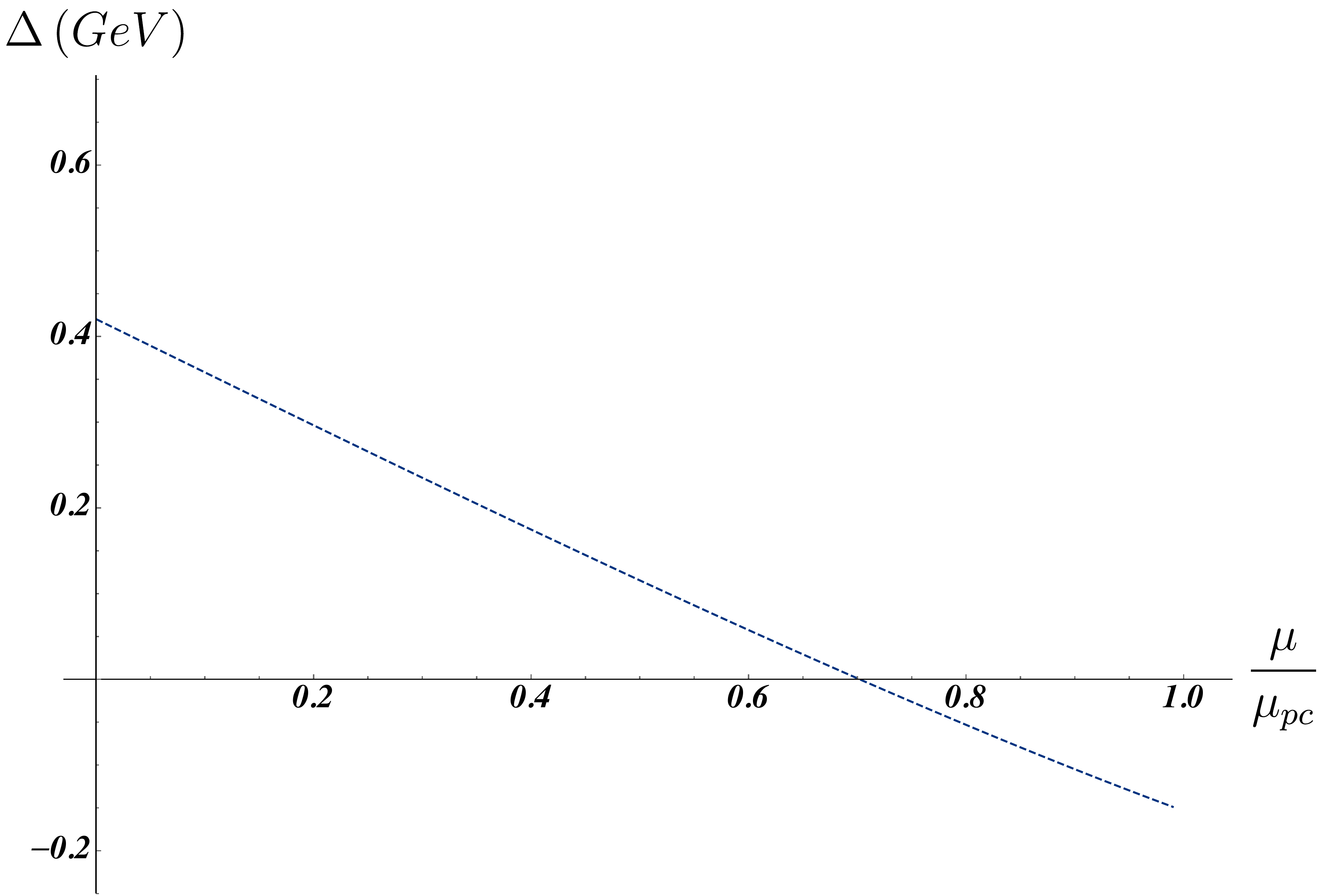}
\caption{{\small The solid curves correspond to the set $L$ with $\r=1.5$, and the dashed ones to $P$ with $\r=3$. Left: The string breaking distances: $\ell_c^{m}$ (in blue) and $\ell_c^{(b')}$ (in red). Right: $\Delta$ as a function of $\mu$. Here $\Delta=M_{\Lambda_c^+}-M_{D^0}$ at $\mu=0$.}}
\label{lc}
\end{figure*}

To make estimates of these distances, we need a value of $\r$. As before, one way would be to use lattice results, but now with the caveat that those are available only at small baryon chemical potential. The estimates of the Debye screening mass yield $2\lesssim\r\lesssim 6$ \cite{a-screen}. In Figure \ref{lc} on the left, we present our results.
We see that the energetic preference of the meson decay mode over the baryon one decreases with chemical potential (baryon density). Moreover, the baryon mode might even become energetically favorable for some values of parameters. Combining this with the natural assumption that the probability for a string rearrangement between the heavy quarks and light quarks coming from the medium also increases with an increase of baryon density, we expect the enhancement of baryon production in strong decays of heavy mesons. This in-medium effect should be more visible at higher densities of baryonic matter. 

\textit{$\Lambda_c^+$ enhancement.--} Recently, the ratio of $\Lambda_c^+$ to $D^0$ production has been measured with the ALICE detector at CERN \cite{alice}. Although the ratios measured in $pp$ and $p\text{Pb}$ collisions turned out to be compatible with uncertainties, the enhanced $\Lambda_c^+$ production was observed in PbPb collisions. Without going deeply into the aspects of hadronization \cite{lee-di}, here we will give a general argument using the minimal free energy principle. Our point is that the effect may be due to formation of a thermalized medium (quark-gluon plasma) in such collisions at higher baryon chemical potential. 

The $\Lambda$-baryons and $D$-mesons are formed from $c$-quark coalescence with light quarks and antiquarks at hadronization. Assuming that the hadrons are at rest in the plasma frame, we can estimate the difference $\Delta$ between the free energies $F_{\Lambda_c^+}$ and $F_{D^0}$. From  \eqref{FQqb} and \eqref{FQqq}, it follows that 
\begin{equation}\label{gap-mu}
\frac{\Delta}{\sqrt{\s}}=
2{\cal Q}\bigl(\g,\m,\r,q,\tfrac{q}{h}\bigr)
+{\cal V}\bigl(\g,v,\tfrac{v}{h}\bigr)
-{\cal Q}\bigl(\g,\m,-\r,\bar q,\tfrac{\bar q}{h}\bigr)
-\tfrac{\mu}{\sqrt{\s}}
\,,
\end{equation}
which we plot in Figure \ref{lc} on the right. Obviously, meson formation becomes less and less energetically favorable as the chemical potential increases. At $\mu\sim 0.7\mu_{pc}$, a transition occurs to a regime where baryon formation is energetically favorable \cite{mupc}. With the ratio $\Lambda_c^+/D^0\sim 1$ measured by ALICE \cite{alice}, it could provide a crude estimate of an upper bound for $\mu$ in PbPb collisions. We expect that temperature effects do not change our main conclusion on the importance of $\mu$. 

\textit{Acknowledgments.--} This research is supported in part by Landau Institute Program 0033-2019-0005. We would like to thank P. de Forcrand, R.R. Metsaev, and P. Weisz for useful discussions. 
\appendix
\renewcommand{\theequation}{A.\arabic{equation}}
\setcounter{equation}{0}
\textit{Appendix: Definitions and some formulas.--} We define the following functions:
\begin{widetext}
\begin{gather}
\varphi(x)=\frac{1}{6}-\frac{1}{6x}\biggl(1+2\sqrt{7x^2-8x+1}
\sin\frac{1}{3}\arcsin\frac{10x^3-75x^2+12x-1}{\bigl(7x^2-8x+1\bigr)^{\frac{3}{2}}}\biggr)
\,,
\label{phi}
\\
{\cal Q}(a,b,c,x,y)=a\Bigl(\sqrt{\pi}\,\text{erfi}(\sqrt{x})-\frac{\ep^{x}}{\sqrt{x}}\Bigr)
+
b(1-y)\sqrt{1+2y}\frac{\ep^{\oh x}}{\sqrt{x}}
+
\frac{1}{3}c\sqrt{\frac{y^3}{x}}
\,,
\qquad
{\cal Q}_0(a,b,x)={\cal Q}(a,b,c,x,0)
\,,
\label{Qb}
\\
{\cal V}(a,x,y)=a\biggl(
\frac{4x(1-y)(1+2y)+6y^2}{(1+4x)(1-y)(1+2y)+6y^2}\frac{\ep^{x}}{\sqrt{x}}
-\sqrt{\pi}\,\text{erfi}(\sqrt{x})
\biggr)
\,,
\qquad
{\cal V}_0(a,x)={\cal V}(a,x,0)
\,,
\label{V}
\end{gather}
where $\text{erfi}(x)$ is the imaginary error function. Then the string tension and $I$ can be written as 
\begin{gather}\label{I}
\sigma(\mu)=\sigma\bigl(1-\varphi \bigr)\sqrt{1+2\varphi}\,\frac{\ep^{h\varphi-1}}{h\varphi}
\,,
\qquad
I=\frac{1}{\sqrt{h\varphi}}
\int_0^1\frac{dx}{x^2}
\biggl(
1+x^2-\ep^{h\varphi x^2}
\biggl[
1-x^4\frac{\bigl(1-\varphi\bigr)^2\bigl(1+2\varphi\bigr)}{\bigl(1-\varphi x^2\bigr)^2\bigl(1+2\varphi x^2\bigr)}\,\ep^{2h\varphi(1-x^2)}
\biggr]^{\oh}
\,\biggr)
\,.
\end{gather}
\end{widetext}



\small
\begin{thebibliography}{99}
\bibitem{strings}
X. Artru, Phys.Rep. {\bf 97}, 147 (1983); N. Isgur and J.E. Paton, Phys.Rev.D {\bf 31}, 2910 (1985). 
\bibitem{diquarks}
R.L. Jaffe, Phys.Rept. {\bf 409} 1, (2005); 
R.F. Lebed, R.E. Mitchell, and E.S. Swanson, Prog.Part.Nucl.Phys. {\bf 93} 143, (2017).
\bibitem{drum}
I.T. Drummond, Phys.Lett.B {\bf 434}, 92 (1998).
\bibitem{bali}
G.S. Bali, H. Neff, T. D\"ussel, T. Lippert, and K. Schilling (SESAM Collaboration), Phys.Rev.D {\bf 71}, 114513 (2005).
\bibitem{bulava}
J. Bulava, B. H\"orz, F. Knechtli, V. Koch, G. Moir, C. Morningstar, and M. Peardon, Phys.Lett.B {\bf 793}, 493 (2019).
\bibitem{V0}
We assume that a string stretched between the heavy quarks is in its ground state.
\bibitem{uaw-book}
The literature on this subject is vast. A good summary with many references may be found in J. Casalderrey-Solana, H. Liu, D. Mateos, K. Rajagopal, and U.A. Wiedemann, Gauge/String Duality, Hot QCD and Heavy Ion Collisions, Cambridge University Press, 2014.
\bibitem{witten}
E. Witten, J. High Energy Phys. {\bf 9807}, 006 (1998).
\bibitem{slant}
For convenience, all the fit parameters are displayed in a mathsf font.
\bibitem{son}
A. Karch, E. Katz, D.T. Son, and M.A. Stephanov, Phys.Rev.D {\bf 74}, 015005 (2006).
\bibitem{az1}
O. Andreev and V.I. Zakharov, Phys.Rev.D {\bf 74}, 025023 (2006).
\bibitem{giannuzzi}
M.V. Carlucci, F. Giannuzzi, G. Nardulli, M. Pellicoro, and S. Stramaglia, Eur.Phys.J. {\bf C57}, 569 (2008).
\bibitem{tach}
Of course, it is quite a crude attempt to describe light sea quarks but a real alternative to probe flavor branes (see, e.g., A. Karch, E. Katz, and N. Weiner, Phys.Rev.Lett. {\bf 90}, 091601 (2003)). Note that in this context, an open string tachyon signals an instability of a fundamental string rather than a D-brane. 
\bibitem{bn-book}
See, e.g., B.M. Barbashov and V.V. Nesterenko, {\it Introduction to the relativistic string theory} (World Scientific, 1990) and references therein.
\bibitem{son}
J. Erlich, E. Katz, D.T. Son, and M.A. Stephanov, Phys.Rev.Lett. {\bf 95}, 261602 (2005).
\bibitem{largeL}
In the string breaking region, $E_{\QQb}(\ell)$ is a linear function. This condition is satisfied for the parameter values we use. 
\bibitem{strb}
O. Andreev, arXiv:2003.09880 [hep-ph].
\bibitem{a-3q}
O. Andreev, Phys.Rev.D {\bf 93}, 105014 (2016).
\bibitem{a-q2}
O. Andreev, Phys.Rev.D {\bf 73}, 107901 (2006).
\bibitem{pdg}
C. Patrignani et al. (Particle Data Group), Chin.Phys.C, {\bf 40}, 100001 (2016). 
\bibitem{PC1}
P. Colangelo, F. Giannuzzi, and S. Nicotri, Phys.Rev.D {\bf 83}, 035015 (2011).
\bibitem{a-screen}
O. Andreev, Phys.Rev.D {\bf 94}, 126003 (2016).
\bibitem{alice}
S. Acharya et al. [ALICE Collaboration], Phys.Lett.B {\bf 793}, 212 (2019).
\bibitem{lee-di}
S.H. Lee, K. Ohnishi, S. Yasui, I-K. Yoo, and C-M. Ko, Phys.Rev.Lett. {\bf 100}, 222301 (2008).
\bibitem{mupc}
A simple estimate gives $435\,\text{MeV}$ for the transition value of $\mu$. It has to be said that there is a big uncertainty associated with the determination of $\r$. 
\end{thebibliography}
\end{document}